\documentclass[useAMS,usenatbib]{mn2e}
\usepackage{natbib}
\usepackage{graphicx}
\usepackage{epstopdf}
\usepackage{amsmath}

\title[Improved magnetogram calibration of SMFT and its comparison with the HMI]{\textbf{Improved} magnetogram calibration of Solar Magnetic Field Telescope and \textbf{its comparison with the Helioseismic and Magnetic Imager}}
\author[X.Y. Bai]{X.Y. Bai$^{1,2}$\thanks{E-mail:
xybai@bao.ac.cn} , Y.Y. Deng$^{1}$, F. Teng$^{1}$, J.T. Su$^{1}$, X.J. Mao$^{3}$, G.P. Wang$^{1}$\\
$^{1}$Key Laboratory of Solar Activity, National Astronomical Observatories, Chinese Academy of
Sciences, 20 Datun Road,\\ Beijing 100012, China\\
$^{2}$Space Environment Prediction Center£¬National Space Science Center, Chinese Academy of Sciences, NO.1 Nanertiao,\\ Beijing 100190, China\\
$^{3}$Astronomy Department, Beijing Normal University, Beijing 100875, China }

\begin{document}

\date{Accepted . Received ; in original form }

\pagerange{\pageref{firstpage}--\pageref{lastpage}} \pubyear{2002}

\maketitle

\label{firstpage}

\begin{abstract}
In this paper, we try to improve the magnetogram calibration method of the Solar Magnetic Field Telescope (SMFT). The improved calibration process fits the observed full Stokes information, using six points on the profile of Fe\, {\i} 5324.18 {\AA} line, and the analytical Stokes profiles under the Milne-Eddington atmosphere model, adopting the Levenberg-Marquardt least-square fitting algorithm. In Comparison with the linear calibration methods, which employs one point, there is large difference in the strength of longitudinal field $B_l$ and tranverse field $B_t$, caused by the non-linear relationship, but the discrepancy is little in the case of inclination and azimuth. We conclude that it is better to deal with the non-linear effects in the calibration of $B_l$ and $B_t$ using six points. Moreover, in comparison with SDO/HMI, SMFT has larger stray light and acquires less magnetic field strength. For vector magnetic fields in two sunspot regions, the magnetic field strength, inclination and azimuth angles between SMFT and HMI are roughly in agrement, with the linear fitted slope of 0.73/0.7, 0.95/1.04 and 0.99/1.1. In the case of pores and quiet regions ($B_l$ $<$ 600 G), the fitted slopes of the longitudinal magnetic field strength are 0.78 and 0.87 respectively.
\end{abstract}


\section{Introduction}

The solar magnetic field, which was first measured according to Zeeman effect slightly more than a century ago \citep{Hale1908}, plays an important role in solar activity and space weather. Two kinds of instruments measuring the magnetic fields have been developed since then. One is the filter-based magnetograph possessing the advantage of imaging observation with high temporal resolutions, such as the Narrowband Filter Imager (NFI) onboard Hinode, the other is the spectral magnetograph, which can acquire the spectral profiles with a wide spectral range, for instance, the Spectro-Polarimeter (SP) onboard Hinode \citep{Tsuneta2008}. In order to combine the advantages of two types of instrument, a new kind of instrument is required, which can obtain image and spectral profile simultaneously (or at least within very short time), e.g., the Helioseismic and Magnetic Imager (HMI) on-board of the Solar Dynamics Observatory (SDO) or the ongoing two-dimensional real-time spectrograph \citep{Deng2009, Schou2012}.

Stokes parameters are first measured by the instruments, i.e., \textit{I, Q, U,} and  $V$, then magnetic fields together with some thermodynamic parameters are inversed through the polarized radiative transfer equation. Up to now, the calibration methods of two kinds of instruments mentioned above have some differences due to their own instrumental characters. The filter-based magnetograph uses one point of the working line profile, thus its calibration is generally under the weak-field assumption that the relationship between the longitudinal field strength $B_{l}$ vs. the Stokes parameters \textit{$V/I$} and the transverse field strength $B_{t}$ vs. $\xi$ is approximately linear \citep{Jeff1989}. Here $\xi$ is defined as $[(Q/I)^{2}+(U/I)^{2}] ^{1/4}$. The key step is to calculate the linear calibration coefficients, which can be obtained with five different methods \citep{Bai2013}. 
However, there is a wide spectral range in the spectral magnetograph, containing enough information of the line formation in the presence of magnetic fields. The key step here is to do the magnetic field inversion in a certain atmosphere model by an appropriate mathematic technique to deal with the polarized spectral data instead of calculating the linear calibration coefficients. In general, two kinds of solutions of the radiative transfer equation, the analytical and numerical solutions, are often employed in the atmosphere model. Up to the present time, the widespread used inversion code is under the assumption of Milne-Eddington atmosphere associated with the analytical solution, developed by \citet{Skumanich1987}. There are also some inversion codes associated with the numerical solution, given by \citet{Socas-Navarro2001}.

Is there a method to measure the magnetic fields with limited spectral points in the new instruments, such as, SDO/HMI or the ongoing two-dimensional real-time spectrograph? \citet{Graham2002} concluded that as least as four measurements analyzed with the inversion technique provide enough information to retrieve the intrinsic magnetic field, and an accuracy better than 10\% is achieved in most cases. \citet{Borrero2011} developed an inversion code named VFISV (Very Fast Inversion of the Stokes Vector), which would routinely analyze pipeline data from SDO/HMI. Recently, \cite{tengf2014} improved the VFISV inversion code by using a smoother interpolation for calculating Voigt function which improves the code performance. \citet{Centeno2014} modified the original VFISV inversion code in order to optimize its operation to the HMI data pipeline and provides the smoothest solution in active regions. 

The Solar Magnetic Field Telescope (SMFT) is a filter-based magnetograph installed at Huairou Solar Observing Station (HSOS), National Astronomical Observatories, Chinese Academies of Sciences \citep{Ai1987}. It has uninterruptedly been observing vector magnetic fields with the photospheric Fe\, {\i} 5324.18 {\AA} line for more than twenty years. Five different methods of the magnetogram calibration under the weak-field assumption have been done since then \citep{Bai2013}. In fact, the relationship between $B_{l}$ vs. Stokes \textit{$V/I$} and $B_{t}$ vs. $\xi$ is non-linear, because the weak-field assumption breaks in the strong magnetic field region, such as in the umbra region where the magnetic saturation occurs. Faraday rotation is another tough problem because of the lack of spectral information \citep{Su2007}. To some extent, these disadvantages can be avoided when more spectral information are available, e.g., measuring six points on the working line profile.

This paper aims to improve the calibration method of SMFT through
measuring six points on the line profile. Furthermore, the
two-dimensional real-time spectrograph has the ability of
acquiring eight positions on the working line profile
simultaneously, and this work provides a base of the magnetic
field inversion. This paper also helps in
understanding the data from HMI if we compare the calibration results
between SMFT and HMI. As it is known that most of the comparisons with
HMI data focused on Line-of-sight magnetogram other than vector
magnetogram \citep{Pietarila2013, Riley2014, Liuy2012, Demidov2011}. Even if there are some comparisons of vector
magnetogram, they are mainly from the data with the spectral magnetograph, which is different from the HMI's \citep{Sainz2011}.

The organization of the paper is as follows. In Section 2,
we briefly describe the observation data and the inversion result.
The magnetogram comparison between the usage of six points and one point on the line
profile with SMFT is presented in Section 3. Section 4 gives a
comparison of the inversion results point by point between SMFT
and HMI.  Conclusions and discussion are arranged at the last
section.

\section{Observations data and inversion result}

\subsection{Intensity profile of SMFT}

SMFT is a refracting telescope, whose aperture is 35 cm. The
passband of the filter of SMFT is 0.15 {\AA}, and the FOV is
$4^{'} \times 3.5^{'}$, with a resolution of $0.242^{''}$
pixel$^{-1}$. Adopting the new observing software
\citep{Shen2013}, the magnetic field sensitivity of $B_l$ is about
10 G, and of $B_t$ about 100 G, if we have a 256-frame integration
in the data used in the paper.

\begin{figure}
\centerline{\includegraphics[width=0.5\textwidth,clip=]{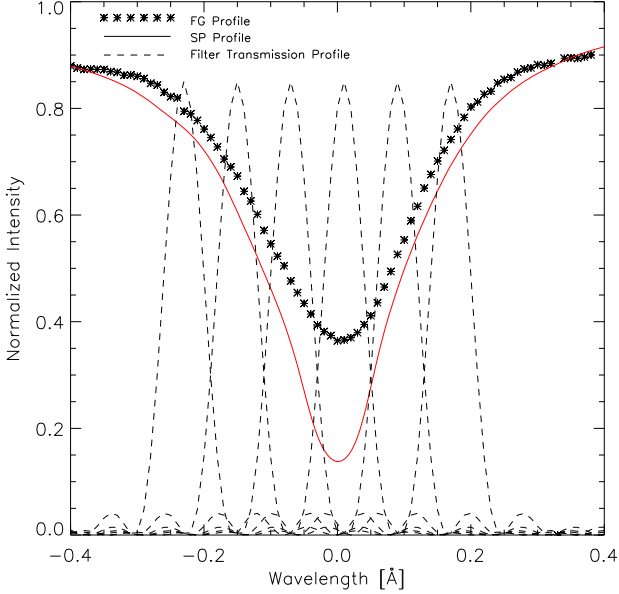}}
\caption{ Scanned spectral profile of SMFT. The asterisk signs represent the scanned profile, and the solid line is the solar disk center spectrum from the Spectral Atlas in Bass2000. The theoretical filter transmission profiles are also plotted with the dashed lines, whose peaks mark the positions of the six points used in the inversion. }
\label{fig-fgp}
\end{figure}

The working wavelength of SMFT is Fe\,{\i} 5324.18 {\AA}. Although
there are some instrumental test in the laboratory during the
maintenance, we obtained the solar spectrum with the spectral scan
method to confirm that all the units of SMFT work well. The
intensity profile is scanned in a quiet region at the center of
solar disk on Nov. 8, 2012. Fig. \ref{fig-fgp} shows the profile
from $-$400m {\AA} to $+$400 m{\AA} around the line center with
the scanned step of 10 m{\AA}. Comparing the scanned profile
(asterisk signs) with the standard solar spectrum (solid line),
they are roughly in agreement. The differences between these two
profiles may be caused by the convolution of the filter
transmission profile or the stray light.

\begin{table}
\begin{center}
\caption[]{Differences in the main properties of SMFT and HMI.}
 \begin{tabular}{c|c|c}
 \hline
  Features & SMFT & HMI\\
  \hline
  Spatial resolution & $ 2\sim3^{''}$ & $1^{''}$ \\
  Spatial sampling  & $0.242^{''}$ & $0.5^{''}$ \\
  FOV  & $240^{''} \times 210^{''}$ & Full disk\\
  Spectral coverage & Fe\, {\i} 5324 {\AA} & Fe\, {\i} 6173 {\AA}\\
  Spectral sampling  & $\sim 80\;m{\AA}$ & $\sim 69\;m{\AA}$\\
  \hline
\end{tabular}
\label{table-1}
\end{center}
\end{table}

\subsection{Polarization data from SMFT and HMI}

\begin{figure}
\centerline{\includegraphics[width=0.5\textwidth,clip=]{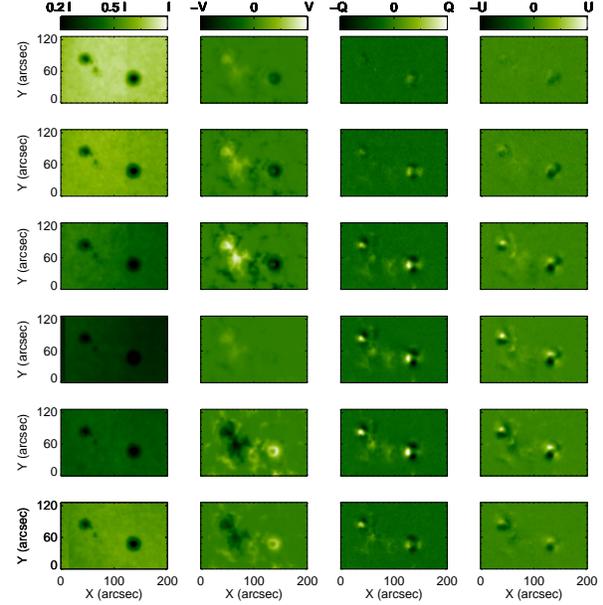}}
\caption{Intensity maps of Stokes \textit{I, V, Q}, and $U$ (from left to right column) in NOAA AR 11611. The six rows from up to down in each column correspond to the selected six wavelength positions at the $-$0.24{\AA}, $-$0.16{\AA}, $-$0.08{\AA}, 0{\AA}, $+$0.08{\AA}, and $+$0.16{\AA} from the line center, respectively.  }
\label{fig-iquv}
\end{figure}

Following the preparatory work, we carried out the Stokes spectral scans observation in the Active Region NOAA 11611 located at N12E13 on Nov. 12, 2012 from 2:40 UT to 3:00 UT. The selected six wavelength positions are at the $-$0.24{\AA}, $-$0.16{\AA}, $-$0.08{\AA}, 0{\AA}, $+$0.08{\AA}, and $+$0.16{\AA} from the line center. The polarization modulator system of SMFT belongs to a type of Stokes definition, i.e. $V_{\pm}=\frac{I \pm V}{2}$ is acquired at the same time with a 256-frame integration. It takes about 20 seconds. Similarly, $Q_{\pm}$ and $U_{\pm}$ are acquired in turn. Then the four Stokes parameters \textit{I, Q, U,} and $V$ are reduced at the six wavelength positions. Fig.\ref{fig-iquv} shows the intensity maps of \textit{I, Q, U,} and $V$. The variations of \textit{I, Q, U,} and $V$ with wavelength are very clear. The value of Stokes $V$ is close to zero at the line center, illustrating that the choice of line center is correct in the data set.

To check the accuracy of the polarization data from SMFT, we downloaded the cotemporal data from HMI, whose observation time was from 02:48 UT to 03:00 UT. They are roughly in agreement (see section 4 for details). The properties and major differences between SMFT and HMI are summarized in Table \ref{table-1}. The resolution of HMI is $0.5^{''}$ pixel$^{-1}$, with the spatial resolution of $1^{''}$. HMI observes the full solar disk while SMFT has a smaller FOV, and therefore a local region extracted out of HMI's full disk is applied in the paper.

\subsection{The inversion result of SMFT and HMI}

The inversion code, developed by \citet{Borrero2011} for SDO/HMI,
is based on the Levenberg-Marquardt least-square fitting algorithm
to inverse magnetic field parameters from the Stokes profiles of
\textit{I, Q, U,} and $V$, adopting the analytical solution under
the assumption of the Milne-Eddington atmosphere model.
\citet{teng2012} extended this code to different spectral lines
and and improved the smoothness of synthetic Voigt functions,
making it possible to inverse SMFT and HMI data.
The formula of the non-linear Least-Square fitting is presented in
Eq. \ref{eq-2}, where \textit{i} indicates the number of observed
wavelength points (\textit{i}=$1\cdot\cdot\cdot6$ in the paper),
and $a_{j}$ refers to a series of model parameters to calculate
profiles analytically, which include: source function ($B_{0}$),
gradient of the source function ($B_{1}$), center to continuum
absorption coefficient ($\eta_{0}$), damping ($a$), Doppler width
of the spectral line ($\triangle\lambda_{D}$), magnetic field
strength ($B$), inclination ($\varphi$), azimuth ($\phi$), and
line-of-sight velocity of the plasma harboring the magnetic field
($V_{los}$).

The atomic parameters of Fe\,{\i} 5324.18 {\AA} are listed in Table \ref{table-2}, with an effective Land\'{e} factor of 1.502. Furthermore, $a$, the damping, is set to 0.5 during the inversion process.


\begin{align}
\chi^2=&\sum_{i}\frac{1}{\sigma_{I}^2}[I_{i}(obs)-I_{i}(a_{j};fit)]^2 \nonumber  \\
&+\sum_{i}\frac{1}{\sigma_{Q}^2}[Q_{i}(obs)-Q_{i}(a_{j};fit)]^2 \nonumber \\
&+\sum_{i}\frac{1}{\sigma_{U}^2}[U_{i}(obs)-U_{i}(a_{j};fit)]^2 \nonumber \\
&+\sum_{i}\frac{1}{\sigma_{V}^2}[V_{i}(obs)-V_{i}(a_{j};fit)]^2 .
\label{eq-2}
\end{align}

\begin{table}
\begin{center}
\caption[]{Atomic parameters for the lower and upper levels of the atomic transition originating the Fe\,{\i}
5324.18 {\AA} spectral line} 
 \begin{tabular}{c|c|c|c}
  \hline
  Level & Electronic config. & J & Land¨¦ factor\\
  \hline
  Upper & $e ^5D$ & 4 & $g_{u}$=1.502 \\
  Lower & $z ^5D^{\circ}$ & 4& $g_{l}$=1.502 \\
  \hline
\end{tabular}
\label{table-2}
\end{center}
\end{table}

The employed transmission profile of the filter \textbf{of SMFT}
is a theoretical one, with the expression of
\begin{align}
T(\lambda)=&\cos^{2}(\pi\frac{\lambda-\lambda_{0}}{0.15})\cos^{2}(\pi\frac{\lambda-\lambda_{0}}{0.30}) \nonumber \\
&\cos^{2}(\pi\frac{\lambda-\lambda_{0}}{0.60}) \cos^{2}(\pi\frac{\lambda-\lambda_{0}}{1.2}),
\end{align}
where $\lambda_{0}$ is the wavelength at the line center. The profiles at the six wavelength positions are plotted in Fig. 1 indicated by dashed lines.

\begin{figure}
\centerline{\includegraphics[width=0.5\textwidth,clip=]{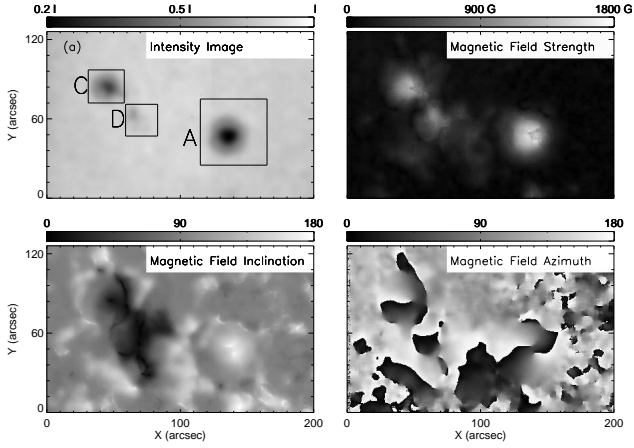}}
\caption{Inversion results of NOAA AR 11611 with SMFT. Top left:
Intensity image; Top right: Magnetic field strength; Bottom left:
Inclination angle of magnetic field vector; Bottom right: Azimuth
angle of magnetic field vector. The squares A, C, and D, in the
top right, are employed to the following quantitative comparison.
} \label{fig-in-smft}
\end{figure}

\begin{figure}
\centerline{\includegraphics[width=0.5\textwidth,clip=]{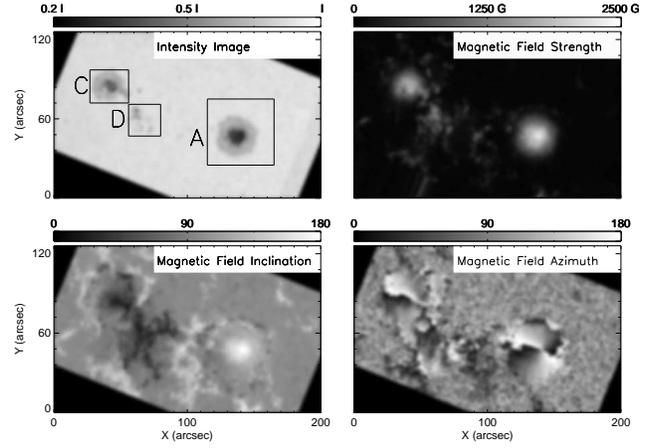}}
\caption{Inversion results of NOAA 11611 with HMI. Top left corner
: Intensity image; Top right corner : Magnetic field strength;
Bottom left corner: Inclination; Bottom right corner: Azimuth. The
squares in the top right corner (a) are the same as Fig. 3.}
\label{fig-in-hmi}
\end{figure}

Here we only consider parts of physical quantities related to the
magnetic properties, which are the total magnetic field strength,
inclination, and azimuth. Fig. \ref{fig-in-smft} shows the
inversion results of NOAA AR 11611 with SMFT.
Similar to Fig. \ref{fig-in-smft}, Fig. \ref{fig-in-hmi}
shows the intensity image and the inversion maps of the total
magnetic field strength, inclination, and azimuth of HMI data.
Three squares, named A, C and D, are chosen to compare magnetic field
parameters of SMFT quantitatively with HMI in section 4. They
contain two sunsopts and several pore regions in Fig. \ref{fig-in-smft}a,
with the size of $50^{''} \times 50^{''}$, $30^{''} \times
25^{''}$, $25^{''} \times 25^{''}$, respectively. In addition, square A is also included in the comparison between using six
points and one point on the working line profile for SMFT, as shown
in the following section.

\section{Comparison of magnetograms calibrated with six and one spectral points on the line profile with SMFT}

The basic equations of magnetogram calibration using one point are:
\begin{align}
B_{l} &= C_{l} (V/I), \nonumber\\
B_{t} &= C_{t} [(Q/I)^{2}+(U/I)^{2}] ^{1/4}=C_{t}\, \xi, \nonumber\\
\phi &= 0.5\tan^{-1}(U/Q).
\end{align}
Here $\phi$ is the azimuth angle, $B_{l}$ and $B_{t}$ represents the longitudinal and transverse components of magnetic fields.
For SMFT, there are five different methods to calculate the calibration coefficients $C_{l}\, $and$\,  C_{t}$ \citep{Bai2013}. In Section 2, magnetic field parameters have been retrieved, based on which we can find the relationship between $B_{l}$ and Stokes $V/I$ , as well as between $B_{t}$ and $\xi$. Moreover, $C_{l}\, $and$\,  C_{t}$ can be derived through a linear fitting on the scatter plots. The selected filter position of $V/I$ is at $-$0.08 {\AA} from the line center, while the position of $Q/I$ and $U/I$ is at the line center. The reason to do so is that the routine observation of SMFT is at the line center for Stokes $Q,\; U$ and at $-$0.075 {\AA} for $V$. Although the position of Stokes $V$ deviates 0.005 {\AA}, the value is far less than the passband (0.15 {\AA}) of SMFT. In this case, the two positions can be treated as the same one.



\begin{figure}
\centerline{\includegraphics[width=0.5\textwidth,clip=]{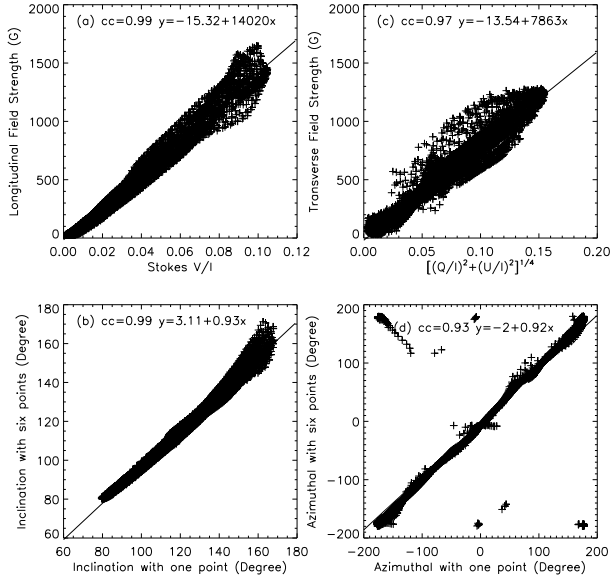}}
\caption{Comparison of individual points of inclination (b) and
azimuth (d) between using six points and one point on the working
line profile in the square A. (a): Scatter plots of the
longitudinal field strength $B_{l}$ vs. \textit{V/I}. (c): Scatter
plots of the transverse magnetic field $B_{t}$ vs. $\xi$. The
solid line represents the linear fit to all the points. The filter
position for the one point case is at $-$0.08 {\AA} and the line
center, corresponding to the longitudinal and transverse magnetic
fields, respectively. } \label{Fig-sixone}
\end{figure}

$B_{l}\, $and$\,  B_{t}$ can be calculated from the following formula if the magnetic field parameters are inferred from six positions on the working line profile:
\begin{align}
B_{l} &= B \cos(\varphi), \nonumber\\
B_{t} &= B \sin(\varphi),
\end{align}
where $\varphi$ is the inclination angle. The scatter plots of
$B_{l}$ vs. Stokes $V/I$ and $B_{t}$ vs. $\xi$ are given in Fig.
\ref{Fig-sixone}a and \ref{Fig-sixone}c, with the data adopted
from the square A in Fig. \ref{fig-in-smft}a. The
correlation coefficient in Fig. \ref{Fig-sixone}a
and \ref{Fig-sixone}c are 0.99 and 0.97 respectively, which show a
striking consistency. In Fig. \ref{Fig-sixone}a, the plot of
$B_{l}$ and Stokes $V/I$ shows a fan-shaped distribution. The
dispersion of $B_{l}$ becomes larger with $V/I$ increasing. The
plot of $B_{t}$ and $\xi$ shows an oval-shaped distribution. When
$\xi$ is at its maximum or minimum, the dispersion of $B_{t}$ is
smaller than that at a median $\xi$. The result reveals
the nonlinear effect between $B_{l}$ vs. Stokes $V/I$ and $B_{t}$
vs. $\xi$, indicating that six points calibration of magnetogram can overcome the disadvantage of linear calibration using
one point.

 $C_{l}$ and $C_{t}$ are obtained from a linear fitting to all
the points, which are
\begin{align}
B_{l} &=-13.918207+14020.727(\pm 367)(V/I), \nonumber\\
B_{t} &=-13.538136+7963.8435(\pm 293) \xi.
\label{eq:lc}
\end{align}
The values in the parentheses are derived from the linear fits and considered to be the errors of the calibration constants.

According to Eq. \ref{eq:lc}, $C_{l}$ and $C_{t}$ are 14020 and 7963, respectively, using six points on the working line profile in the magnetic field inversion. Table \ref{table-3} displays $C_{l}$ and $C_{t}$ obtained from different methods. The value of $C_{l}$ is slightly larger than those of the empirical calibration \citep{Wangj1996}, the observational calibration \citep{Wang1996}, and the nonlinear least-squares fitting calibration \cite{Su2004}. The value of $C_{t}$ is between the empirical and the nonlinear least-squares fitting calibration. After comparing seventeen magnetograms between SP/Hinode and SMFT co-temporally, \cite{Wang2009} concluded that $C_{l}$ and $C_{t}$ are, as it is, systemically larger than those from \cite{Su2004}. Our result is consistent with their conclusion. The reason that the $C_{l}$ and $C_{t}$ are lower is possibly that \citet{Su2004} did not include the umbra.

\begin{table}
\begin{center}
\caption[]{$C_{l}$ and $C_{t}$ calculated with different methods. The filter position is at $-$0.075 {\AA} and at the line center for $B_{l}$ and $B_{t}$, respectively.}
 \begin{tabular}{c|c|c}
 \hline
  Method & Empirical calibration & Observational calibration \\
 \hline
 $C_{l}$ & 10000 & 9600  \\
 $C_{t}$ & 9730 &  \\
  \hline
  Method & Nonlinear least-squares & calibration with six points \\ & fitting calibration &\\
  \hline
  $C_{l}$ & 8381 & 14021\\
  $C_{t}$ & 6790 & 7964\\
  \hline
\end{tabular}
\label{table-3}
\end{center}
\end{table}

Fig. \ref{Fig-sixone} also shows the comparison results of the inclination and azimuth angle between using six points and one point on the working line profile. The inclination angle between the two methods has a strong correlation, whose correlation coefficient is 0.99 and the slope 0.93 in Fig. \ref{Fig-sixone}b. Regarding the magnetic field azimuth in Fig. \ref{Fig-sixone}d, the correlation coefficient is 0.93 and the slope of the linear fit of the plots is 0.92. The values of the two methods are still considerably consistent, except that the values using six points are a little bit lower. Here the azimuth values go from $-$180$^{\circ}$ to $+$180$^{\circ}$, after the correction of 180 degree azimuth ambiguity by means of the potential method. The dots concentrated in the corners diagonally other than the solid line, may result from the failed correction of 180 degree azimuth ambiguity.

%
%


\section{Magnetogram Comparison between SMFT and HMI}

\begin{figure}
\centerline{\includegraphics[width=0.5\textwidth,clip=]{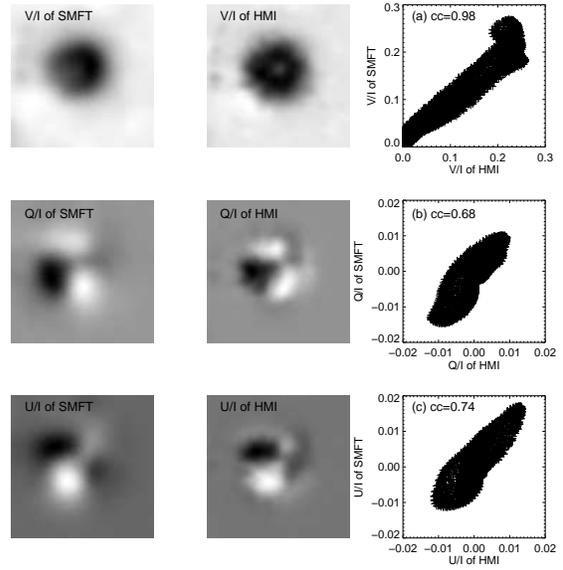}}
\caption{Comparisons of \textit{V/I, Q/I}, and \textit{U/I} between
SMFT and HMI. The left and middle column
represent the data from SMFT and HMI respectively. The right one from up to down represents the
point by point comparison between SMFT and HMI in \textit{V/I, Q/I}, and \textit{U/I}.    } \label{Fig-quv}
\end{figure}

\begin{figure}
\centerline{\includegraphics[width=0.5\textwidth,clip=]{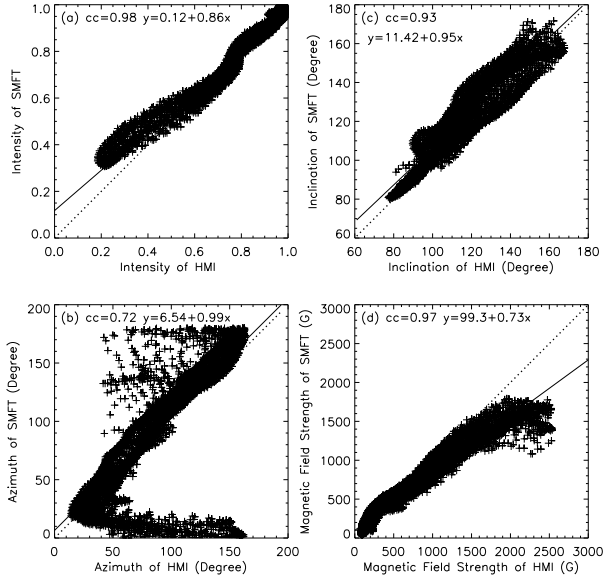}}
\caption{Comparison of individual points of SMFT and HMI in square
A. (a) intensity; (b) azimuth; (c) inclination; (d) total magnetic
field strength. The dotted lines in the four sub-images indicate
perfect correlation between the two axes while the solid lines are
the linear fittings of the scatter points.} \label{fig-con-hmi}
\end{figure}

Due to the pixel resolution difference between HMI and SMFT, a pixel-to-pixel comparison of the inversion results between the two instruments requires an alignment as a first step. Selecting and cutting out the region of interest from HMI maps, we applied a rotation and a pixel scaling to align the intensity maps. Then the co-alignment parameters derived from the previous procedure were applied to the maps of magnetic field strength, inclination, and azimuth. We adopted the CONGRID function in IDL to reform the SMFT data achieving the resolution of $0.5^{''}$ pixel$^{-1}$. Then the SMOOTH function was used for SMFT data to the spatial resolution of $3^{''}$, which is the typical value of most ground-based telescopes without adaptive optics. In the case of HMI data, a two-dimensional Gaussian function is convolved to reduce its spatial resolution to be comparable with SMFT's data. The FWHM of the Gaussian function is 6 pixels, corresponding to the spatial resolution of $3^{''}$. The left and middle column in Fig. \ref{Fig-quv} gives the maps of \textit{V/I, Q/I} and \textit{U/I} from SMFT and HMI using the data in the square A defined in Fig. \ref{fig-in-smft}a. The filter position of Stokes \textit{Q/I} and \textit{U/I} in HMI data is at the third point from the blue wing while that of \textbf{V/I} is at the second point. In the case of the SMFT's data, the filter position is the same as that used in section 3 using one point. The morphological comparison also shows a high degree of consistency, but we can also see some slight difference. The last column in Fig. \ref{Fig-quv} shows the point by point comparison of \textit{V/I, Q/I} and \textit{U/I} between SMFT and HMI, with the corresponding linear correlation coefficient of 0.98, 0.68, 0.74, respectively. The correlation coefficient of \textit{V/I} is larger than that of \textit{Q/I} and \textit{U/I}, due to the weaker linear polarization signals than that of circular polarization. The smaller correlation coefficient in \textit{Q/I} and \textit{U/I} may cause some deviations in the azimuth, shown in Fig. \ref{fig-con-hmi}b.

Through morphological comparisons of Fig. \ref{fig-in-smft} and Fig. \ref{fig-in-hmi}, it can be found that the intensity image of HMI is slightly clear than that of SMFT even if we have convolved a Gauss function in the data recorded by HMI, implying that there are still some differences caused by the effect of atmosphere, which is not eliminated in the observation of SMFT. The maps of magnetic field strength, inclination, and azimuth from SMFT are roughly in agreement with HMI data. Comparisons of individual points in the square \textbf{A}, defined in Fig. \ref{fig-in-smft}a, are shown in Fig. \ref{fig-con-hmi}. Fig. \ref{fig-con-hmi}a is the scatter plots of the intensity normalized to the maximum value. The dotted line represents the 1:1 exact corresponding, and any deviation from that is due to misalignments, non-simultaneity, different sensitivity of the spectral lines to temperature and various other instrumental effects and so on. The linear correlation coefficient of the intensity between SMFT and HMI is 0.98, which is a strong correlation. The solid line is the linear fit of the scatter plots, whose slope is 0.86. The stray light of SMFT is slightly larger than HMI, especially in the umbra with low intensity, where the plots show a larger deviation from the dotted line.

From Fig. \ref{fig-con-hmi}b and \ref{fig-con-hmi}c, most of the magnetic field azimuth and inclination are along the dotted lines with small deviation, indicating that the two parameters of SMFT and HMI have good correspondence. The slopes of the linear fit (solid line) of the scatter plots in Fig. \ref{fig-con-hmi}b and Fig. \ref{fig-con-hmi}c are 0.99 and 0.95, and the corresponding linear correlation coefficients are 0.72 and 0.93, respectively. Here only the areas greater than 300G are employed in the plots of Fig. \ref{fig-con-hmi}b and \ref{fig-con-hmi}c to avoid the noise. Even so, there is still a certain amount of scatter, especially in the comparison of azimuth angle, inferred from Stokes \textit{Q} and \textit{U}, whose signals are much weaker than Stokes $V$ (see Fig. \ref{Fig-quv}). The linear correlation coefficient of magnetic field strength from the two instruments is 0.97, shown in Fig. \ref{fig-con-hmi}d. The slope of the linear fit is 0.73, revealing the magnetic field strength of SMFT is slightly less than that of HMI, particularly in a stronger field strength region such as in an umbra. This phenomenon may be caused by more stray light in SMFT, which can reduce the magnitude of Stokes $V$ \citep{Su2007}.

\begin{figure}
\centerline{\includegraphics[width=0.5\textwidth,clip=]{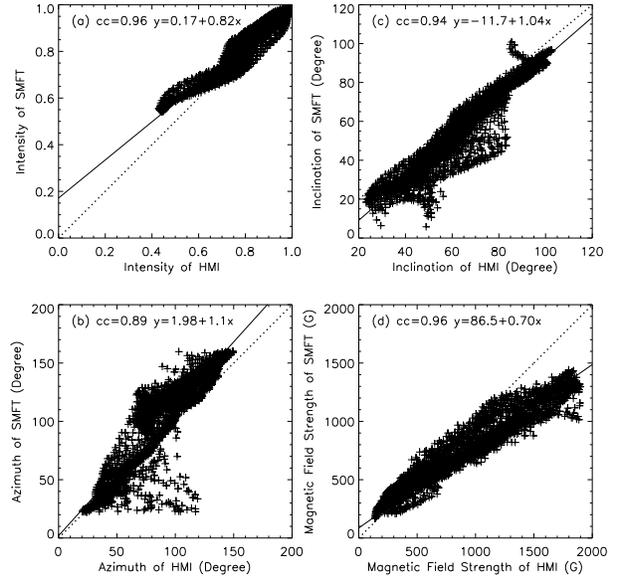}}
\caption{ Same as Fig. \ref{fig-con-hmi}, but for the square C.}
\label{fig-con-hmi2}
\end{figure}

\begin{figure}
\centerline{\includegraphics[width=0.5\textwidth,clip=]{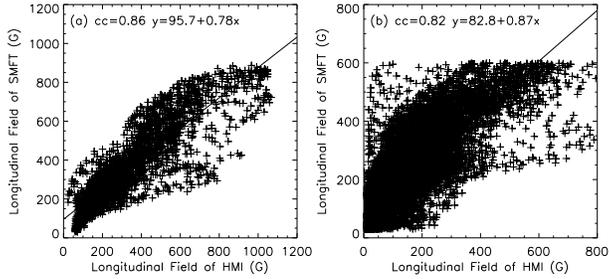}}
\caption{Comparison of individual points of longitudinal field
strength $B_{l}$ between SMFT and HMI.  (a): the result in the
square D. (b): the result when $B_{l}$ less than 600 G. The solid
lines are the linear fittings of the scatter points.}
\label{fig-con-hmi3}
\end{figure}

Fig. \ref{fig-con-hmi2} gives the comparison result of
SMFT and HMI in a smaller sunspot with lower magnetic field
strength in the square C. The comparison of intensity in Fig.
\ref{fig-con-hmi2}a, whose linear correlation coefficient and the
slope of linear fit are 0.96 and 0.82, also shows that SMFT has a
larger stray light. If we focus on the inclination, azimuth and
magnetic field strength, the linear correlation coefficients are
0.94, 0.89 and 0.96 respectively, with 1.04, 1.1 and 0.7 as the corresponding slopes of linear fit. The slopes of the linear fit
of the inclination, azimuth and magnetic field strength are
comparable with that of the larger sunspot in square A. The magnetic field strength
from SMFT in square C is still less than that from HMI.

The comparison of longitudinal field strength between SMFT
and HMI is shown in Fig. \ref{fig-con-hmi3}a for several pore
regions in the square D, having no penumbra and with the magnetic field
strength less than that of a sunspot. As the transverse magnetic field strength is weak, it is not used for
comparison. The linear correlation coefficient of the scatter plot is
0.86, which is slightly lower than that in one of the sunspot regions
in the square A and C. The possible reason is that the area of a pore
is smaller than that of a sunspot, so it is much easily
influenced by the effect of the atmosphere in the data of SMFT and
the alignment between SMFT and HMI data. The longitudinal magnetic field strength from SMFT is less than that from HMI and the slope of the linear
fit of the plots is 0.78.  

In the case of quiet regions with lower magnetic field strength, such as
network regions etc., we also compare its longitudinal field
strength between SMFT and HMI, as displayed in Fig.
\ref{fig-con-hmi3}b. The threshold between 30 G and 600 G in the
data of SMFT in Fig. \ref{fig-in-smft} are selected for comparison. The linear
correlation coefficient is similar to that in the pore regions
(square D), with the value of 0.82. Moreover, the slope of the
linear fits is 0.87 revealing that the longitudinal magnetic field strength
from SMFT is still less than that from HMI. Interestingly, the
slope is larger than that in the pore regions (square D), showing that the
slopes of linear fit between SMFT and HMI varies with longitudinal magnetic
field strength. As the longitudinal magnetic field strength gets lower,
the calibration result between SMFT and HMI appears much closer. 

\section{Conclusion and Discussion}
In this paper, we improve the calibration method of SMFT using six points on the line profile and demonstrate the difference in the magnetic field parameters as compared with other calibration methods using one point. Then we compare the calibrated magnetogram from SMFT with that from HMI in different regions, such as sunspots and pores. The main conclusions are summarized as follows.

\begin{enumerate}
\item With the inversion algorithm using six points on the line profile, we obtained the total magnetic field strength, inclination and azimuth observed by SMFT. At the position of -0.08 {\AA} from the line center and at the line center, the relationship of $B_{l}$ and Stokes $V/I$ shows a fan-shaped distribution while the one of $B_{t}$ and $\xi$ shows an oval-shaped distribution as observed, which are non-linear. The main difference of $B_l$ and $B_t$ in magnetogram calibration between using six points and one point on the working line profile concentrates on the region where the non-linear effect occurs, but the difference of the inclination and azimuth are not very significant (Historically, an interpolation method was employed based on linear calibration to deal with the non-linear effect in HSOS). Comparing the linear calibration coefficients calculated from six points on the working line with the other existent calibration methods on SMFT, the value of $C_{l}$ is larger than the results of the empirical, the observational, and the nonlinear least-squares fitting calibration. The value of $C_{t}$ is between empirical and nonlinear least-squares fitting calibration.

\item The magnetic field parameters of SMFT is approximately in consistence with that of HMI, if we adopt the improved calibration method using six points on the line profile. SMFT has larger stray light from the comparison of intensity, especially in the umbra, and it acquires less magnetic field strength, maybe caused by its larger stray light. The inclination and azimuth angles of the two instruments in two sunspot regions are roughly in agrement, with the linear fitted slope of 0.95/1.04 and 0.99/1.1. The linear correlation coefficient in lower magnetic field strength regions such as pores and networks is less than that in the sunspots regions, probably resulting from the different response of atmosphere effect and the alignment process. Nevertheless, in the case of different magnetic field strength, the fitted slopes are different. For sunspot regions, the slope of total magnetic field strength between SMFT and HMI is about 0.7. In the case of pore regions, the slope of longitudinal magnetic field strength $B_{l}$ is 0.78. Finally the value rises to 0.87 in the case of quiet regions when $B_{l}$ is less than 600 G.
\end{enumerate}

\citet{Wang1992} compared the vector magnetograms obtained at Big Bear Solar Observatory (BBSO), Mees Solar Observatory (MSO) and Huairou Solar Observing Station (HSOS). The general conclusion is that the longitudinal
fields agree better than transversal fields among magnetograms from three observatories, and the agreement of vector fields is better between BBSO and HSOS than between BBSO and MSO. \citet{Zhang2003} analyzed the vector magnetograms from HSOS, MSO and National Astronomical Observatory of Japan, and found that there is a basic agreement on the transversal fields among these magnetographs. In this paper, we have a comparison of magnetic field strength, inclination and azimuth between SMFT and HMI, and conclude that these two instruments have good consistency. The effect of the 12-hour and 24-hour periodicities in strong HMI magnetic fields is not considered in the comparison \citep{Liuy2012}. The differences in the quasi-simultaneous original data from different spectral lines, corresponding to various layers in the solar atmosphere with diverse physical and thermodynamical properties, maybe the cause of the discrepancies in magnetic field strength, inclination, and azimuth, and the inversion processes may also add some on it. In addition, instrumental stray light, polarization accuracy, and spatial resolution are not identical between SMFT and HMI. Finally, the alignment process (rotation and pixel scaling) requires an interpolation, which may introduce differences, too.

\section*{Acknowledgments}
We would like to thank all of the optical and electrical engineers at HSOS. We also acknowledge the free data usage
policy of the SDO/HMI. This work is supported by the grants: KJCX2-EW-T07, 2011CB-811401, 11373040, 11178005, 10878004, 11221063, 11203036, 11273034, 11373044, 11303052, 11303048, the young researcher grant
of national astronomical observatories, and the specialized
research fund for state key laboratories.

\bibliographystyle{mn2e}
\bibliography{5324article}
\end{document}